\RequirePackage{amsmath}
\documentclass[runningheads]{llncs}
\usepackage{graphicx}
\usepackage[usenames,dvipsnames]{xcolor}
\usepackage{tikz}
\usepackage{amssymb}
\usepackage{ebproof}
\usepackage{newunicodechar}
\usepackage{xspace}
\usepackage{multirow}
\usepackage[hidelinks]{hyperref}
\usepackage{enumerate}
\usepackage{comment}
\usepackage{array}

\usepackage{enumitem}
\setitemize{noitemsep,topsep=2pt,parsep=2pt,partopsep=2pt}

\newcommand{\chypo}[1]{
  \hypo{\colorbox{Apricot}{$#1$}}
}

\def\BaseUrl{https://hferee.github.io/UIML}
\newcommand{\coqdoc}[1]{\href{\BaseUrl/#1}{\raisebox{-.9mm}{\includegraphics[height=1em]{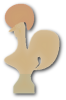}}}}

\newcommand{\goodbox}{\hspace{-.6ex}\text{
    \tikz[baseline=-.6ex, rounded corners=.01ex, line width=.1ex]
    {\draw (-.6ex,-.6ex) rectangle (.6ex,.6ex);}}\kern.2ex}
\renewcommand{\Box}{\goodbox}

\newcommand{\goodcrossedbox}{\hspace{-.6ex}\text{
    \tikz[baseline=-.6ex, rounded corners=.01ex, line width=.1ex]
    {\draw (-.6ex,-.6ex) rectangle (.6ex,.6ex);
      \draw (-.6ex,.6ex) -- (.6ex,-.6ex);
      \draw (-.6ex,-.6ex) -- (.6ex,.6ex);}}\kern.2ex}

\newcommand{\gooddiamond}{\hspace{-.6ex}\text{
    \tikz[baseline=-.6ex, rounded corners=.01ex, rotate=45, line width=.1ex]
    {\draw (-.5ex,-.5ex) rectangle (.5ex,.5ex);}}\kern.2ex}
\renewcommand{\Diamond}{\gooddiamond}

\renewcommand{\phi}{\varphi}

\newcommand{\sys}[1]{\mathsf{#1}}
\newcommand{\IL}{\sys{IL}} 
\newcommand{\K}{\sys{K}}

\newcommand{\KS}{\sys{KS}}
\newcommand{\GL}{\sys{GL}}
\newcommand{\iGL}{\sys{iGL}}
\newcommand{\GLS}{\sys{GLS}}
\newcommand{\iSL}{\sys{iSL}}
\newcommand{\GfouriSLt}{\sys{G4iSLt}}
\newcommand{\GfouriP}{\sys{G4iP}}

\newcommand{\sbf}[1]{\text{Sub}(#1)}

\newcommand{\pvf}[1]{\text{Vars}\,(#1)}
\newcommand{\pvs}[1]{\text{Vars}\,(#1)}

\renewcommand{\rule}[1]{(\text{#1})}
\newcommand{\Ra}{\Rightarrow}
\newcommand{\GLR}{\rule{GLR}}
\newcommand{\IdP}{\rule{IdP}}
\newcommand{\IdB}{\rule{IdB}}
\newcommand{\KR}{\rule{KR}}
\newcommand{\botL}{\rule{$\bot$L}}
\newcommand{\impR}{\rule{$\rightarrow$R}}
\newcommand{\impL}{\rule{$\rightarrow$L}}

\newcommand{\impocc}[1]{imp(#1)} 
\newcommand{\symnumusablebox}[0]{\beta}
\newcommand{\numusablebox}[1]{\symnumusablebox(#1)}
\newcommand{\symGLseqmeas}[0]{\Theta}
\newcommand{\GLseqmeas}[1]{\symGLseqmeas(#1)}
\newcommand{\lexnat}{<\!\!<} 

\newcommand{\symUIK}{\mathsf{A}_{p}^{\scriptscriptstyle{\mathsf{K}}}} 
\newcommand{\UIK}[2]{\symUIK(#2)} 

\newcommand{\symUIGL}{\mathsf{A}_{p}^{\scriptscriptstyle{\mathsf{GL}}}} 
\newcommand{\UIGL}[2]{\symUIGL(#2)} 
\newcommand{\symN}{\mathsf{N}} 
\newcommand{\N}[1]{\symN_{p}(#1)} 

\newcommand{\fc}[1]{\overline{#1}}

\newcommand{\ApiSL}[1]{\mathsf{A}_p^\iSL(#1)}
\newcommand{\EpiSL}[1]{\mathsf{E}_p^\iSL(#1)}
\newcommand{\symApiSL}{\mathsf{A}_p^{\iSL}}
\newcommand{\symEpiSL}{\mathsf{E}_p^{\iSL}}

\newcommand{\canopy}[1]{\text{Can}(#1)}
\newcommand{\GLRprems}[1]{\text{GP}(#1)}
\newcommand{\KRprems}[1]{\text{KP}(#1)}

\newcommand{\openboxes}{\Box^{\scriptscriptstyle{-1}}}
\newcommand{\disjunion}{,}
\newcommand{\boxR}{\ensuremath{\square_R}}

\newcommand{\weight}{\texttt{weight}}

\newcommand{\Ap}[1]{\mathsf{A}_{p}(#1)}
\newcommand{\Ep}[1]{\mathsf{E}_{p}(#1)}
\newcommand{\To}{\Rightarrow}

\newcommand{\isdef}{:=}

\newcommand{\ApIPC}{\mathsf{A_p^\IL}}
\newcommand{\EpIPC}{\mathsf{E_p^\IL}}
\newcommand{\callAp}[1]{\mathsf{\mathcal{A}_p}(#1)}
\newcommand{\callEp}[1]{\mathsf{\mathcal{E}_p}(#1)}

\newcommand{\V}{\cal V}

\newunicodechar{⊙}{\ensuremath{\odot}}
\newunicodechar{⊗}{\ensuremath{\otimes}}
\newunicodechar{□}{\ensuremath{\Box}}
\newunicodechar{Δ}{\ensuremath{\Delta}}
\newunicodechar{δ}{\ensuremath{\delta}}
\newunicodechar{θ}{\ensuremath{\theta}}
\newunicodechar{ϕ}{\ensuremath{\phi}}
\newunicodechar{ψ}{\ensuremath{\psi}}
\newunicodechar{Γ}{\ensuremath{\Gamma}}
\newunicodechar{≺}{\ensuremath{\prec}}
\newunicodechar{∖}{\ensuremath{\setminus}}
\newunicodechar{∧}{\ensuremath{\wedge}}
\newunicodechar{∨}{\ensuremath{\vee}}
\newunicodechar{⊼}{\ensuremath{\barwedge}}
\newunicodechar{⊻}{\ensuremath{\veebar}}
\newunicodechar{₁}{\ensuremath{_1}}
\newunicodechar{₂}{\ensuremath{_2}}
\newunicodechar{₃}{\ensuremath{_3}}
\newunicodechar{⊥}{\ensuremath{\bot}}
\newunicodechar{⊤}{\ensuremath{\top}}
\newunicodechar{∈}{\ensuremath{\in}}
\newunicodechar{⇁}{\ensuremath{\rightharpoondown}}
\newunicodechar{∀}{\ensuremath{\forall}}
\newunicodechar{∃}{\ensuremath{\exists}}
\newunicodechar{λ}{\ensuremath{\lambda}}
\newunicodechar{⊢}{\ensuremath{\vdash}}
\newunicodechar{≡}{\ensuremath{\equiv}}
\newunicodechar{⋀}{\ensuremath{\bigwedge}}
\newunicodechar{⋁}{\ensuremath{\bigvee}}
\newunicodechar{∅}{\ensuremath{\emptyset}}
\newunicodechar{⊎}{\ensuremath{\uplus}}

\usepackage{bbding}

\begin{document}

\title{Mechanised uniform interpolation for modal logics K, GL, and iSL}
\author{Hugo Férée\inst{1}\orcidID{0000-0003-3103-5612}\and  
Iris van der Giessen\inst{2}\orcidID{0009-0008-4908-2496} \and 
Sam van Gool\inst{1}\Envelope\orcidID{0000-0002-6360-6363} \and
Ian Shillito\inst{3}\orcidID{0009-0009-1529-2679}}
\authorrunning{H. Féréé et al.}

\institute{Université Paris Cité, CNRS, IRIF, F-75013, Paris, France \and University of Birmingham, Birmingham, UK
\and Australian National University, Canberra, Australia}

\maketitle
\setcounter{footnote}{0}
\begin{abstract}
The uniform interpolation property in a given logic can be understood as the
definability of propositional quantifiers. We mechanise the computation of these
quantifiers and prove  correctness in the Coq proof assistant for three modal
logics, namely: (1) the modal logic K, for which a pen-and-paper proof exists;
(2) Gödel-Löb logic GL, for which our formalisation clarifies an important point
in an existing, but incomplete, sequent-style proof; and (3) intuitionistic
strong Löb logic iSL, for which this is the first proof-theoretic construction
of uniform interpolants. Our work also yields verified programs that allow one
to compute the propositional quantifiers on any formula in this logic.
\end{abstract}

\section{Introduction}

Uniform interpolation is a strong form of interpolation, which says that
propositional quantifiers can be defined inside the logic. More precisely, a left uniform interpolant
of a formula $\phi$ with respect to a variable $p$ is a $p$-free formula, denoted $\forall p \phi$,
which entails $\phi$, and is a consequence of any $p$-free formula that entails $\phi$. The dual
notion is that of a right uniform interpolant, denoted $\exists p \phi$, and a logic is said to have
uniform interpolation if both left and right uniform interpolants exist for any formula. Said otherwise, uniform interpolation means that  for any $\phi$ and $p$, the logic has a strongest formula without $p$ that implies $\phi$, and a weakest formula without $p$ that is implied by $\phi$.

The uniform interpolation property was
first established for intuitionistic propositional logic $\IL$ by Pitts~\cite{Pit92}, and then for
a number of modal logics, including basic modal logic $\K$ and Gödel-Löb provability logic $\GL$~\cite{Shavrukov,Vis1996,GhiZaw2002}.
Since then, uniform interpolation has been
shown to hold in various modal fixpoint logics~\cite{AgoHol2000,MarSeiVen2015} and substructural logics~\cite{AliDerOno2014}, and
connections have been developed with
description logic~\cite{GhiLutWol2006}, proof theory~\cite{Iem2019,Gie22}, model theory~\cite{GhiZaw2002,Koc2023},
and universal algebra~\cite{GooMetTsi2017,KowMet2019}.

Existing
proof methods for uniform interpolation can be divided, roughly, into two strands: one is syntactic
and relies on the existence of a well-behaved sequent calculus for the logic (see e.g.~\cite{Iem2019}),
the other is semantic and uses Kripke models to establish definability of bisimulation quantifiers (see e.g.~\cite{GhiZaw2002}).
An advantage of the syntactic method over the semantic one
is that, at least in theory, it provides better bounds on the complexity
of computing uniform interpolants. In practice, however, it is not feasible to compute
uniform interpolants by hand, as the calculations quickly become complex even on small examples.
The algorithms for computing uniform interpolants are often intricate,
and it is a non-trivial task to implement them correctly.
The first- and third-named author recently developed the first verified implementation of
Pitts' algorithm for computing uniform interpolants in the case of $\IL$, using The
Coq Proof Assistant in order to formally prove the correctness of the implementation~\cite{Fer23}.

In this article, we provide
mechanised proofs of the uniform interpolation property for the classical modal logics
$\K$ and $\GL$ and for an intuitionistic version of strong Löb logic, $\iSL$.
Of these three contributions, we discuss the first one in Section~\ref{sec:K},
which serves as a warm-up for what follows. The formalisation of uniform interpolation
for $\GL$  starts from a sequent-style
proof of this theorem~\cite{Bil22}.
During our work on formalising this proof in Coq, we uncovered an incompleteness in it,
and our formalisation contains a corrected version of the construction of~\cite{Bil22},
as we will explain further  in Section~\ref{sec:GL}.
Finally, the uniform interpolation result for $\iSL$ is new to this paper, and resolves an open question
of \cite{GieIem2020}.
({T. Litak and A. Visser have shared a draft paper with us in which they obtain a different, semantic, proof of the same result, available in preprint \cite{LitakVisser24}.})
The proof we give extends the syntactic method of Pitts,
while taking
advantage both of the robustness of the earlier Coq formalisation for the case of $\IL$, and of
a recently developed sequent calculus for $\iSL$~\cite{Shi23}.

All definitions and proofs that we describe in this paper are implemented in the constructive setting
of the Coq proof assistant; the code is available online at \url{https://github.com/hferee/UIML}.
In particular, this means that the definitions of the uniform interpolants for the three logics at hand here
are effective, which allows us to extract from the Coq implementation an OCaml program that can
generate interpolants from input formulas.
Throughout the paper, links to an online-readable version
of the Coq proofs are given by a clickable symbol \includegraphics[height=1em]{coql.png}.
Finally, a demonstration webpage is available at \url{https://hferee.github.io/UIML/demo.html} where the uniform interpolants for each logic can be computed.

\section{Sequent calculi and uniform interpolation}

In this section, we recall some standard notions that we need in this paper, pertaining to the classical modal logics $\K$ and $\GL$, and intuitionistic modal logic $\iSL$. We mostly follow the same notations as in \cite[Ch.~1]{Gie22}, and we refer the reader to that chapter for more details.

It will be convenient to use a more economical language for the classical setting than for the intuitionistic setting, so we define the precise syntax in some detail now. Both languages contain \textit{boolean constant} $\bot$, \textit{connective}~$\to$,  \textit{modality} $\Box$ and a set $\V$ of countably many \emph{(propositional) variables}, denoted  $p,q, \dots$ . 

In the \textit{classical modal language} we use the following standard classical constructors, $\neg$, $\vee$, $\wedge$, and
$\Diamond$, which should be read as abbreviations: $\neg \phi := \phi \to
\bot$, $\phi \vee \psi := (\phi \to \bot) \to \psi$, $\phi \wedge \psi := (\phi \to (\psi \to \bot)) \to \bot$, and $\Diamond \phi :=
\Box (\phi \to \bot) \to \bot$.
The \textit{intuitionistic modal language}, instead contains the \textit{connectives}~$\wedge$, $\vee$ (no $\Diamond$) ; only $\neg$ and $\top$ are abbreviations: $\neg \phi := \phi \to \bot$, $\top := \neg \bot$.
In both the classical and intuitionistic setting, we denote modal formulas by lowercase Greek letters $\phi, \psi, \ldots$ and we write $\pvf{\phi}$ to denote the set of all propositional variables occurring as subformulas in the formula $\phi$.

We briefly recall the axiomatisation of logics $\K$, $\GL$, and $\iSL$. The
logics $\K$ and $\GL$ are defined over the considered classical modal language
and $\iSL$ over the intuitionistic modal language. To do so, we recall three
axioms:
\begin{itemize}
    \item the \textit{normal axiom} $(\mathsf{k}) \  \Box (p \to q)  \to \Box p \to \Box q$, 
    \item the \textit{G\"{o}del-L\"{o}b axiom} $(\mathsf{gl}) \ \Box (\Box p \to p)
        \to \Box p$, \quad and
    \item the \textit{strong L\"{o}b axiom} $(\mathsf{sl}) \   (\Box p \to p) \to p$. 
\end{itemize}
Also recall the rules  \textit{modus ponens} (from $\phi$ and $\phi \to \psi$
infer  $\psi$), \textit{necessitation} (from $\phi$ infer $\Box \phi$), and
\textit{substitution} (from $\phi$ infer $\sigma \phi$, for any uniform
substitution $\sigma$)). Now, logic $\K$ is defined by
the classical propositional tautologies, axiom~$\sf k$, and the rules modus
ponens, necessitation, and substitution. The logic $\GL$ is the extension of $\K$
by the axiom $\sf{gl}$. Furthermore, intuitionistic propositional logic $\IL$ is
defined by the intuitionistic tautologies, and the rules modus ponens,  
necessitation, and substitution; intuitionistic modal logic $\iSL$ is the extension of $\IL$
with axioms~$\sf k$ and~$\sf{sl}$.

\subsection{Sequent calculi}
A \emph{sequent} is a pair of finite multisets of formulas $\Gamma$ and $\Delta$,
which we denote by $\Gamma \To \Delta$. In the intuitionistic case, $\Delta$
will necessarily be a singleton. A sequent $\Gamma \To \Delta$ is
\textit{empty}, if $\Gamma$ and $\Delta$ are empty multisets. Given two multisets
$\Gamma$ and $\Delta$, we
write $\Gamma,\Delta$ for the multiset addition of $\Gamma$ and $\Delta$, and,
when $\phi$ is a formula, we write $\Gamma, \phi$ as notation for $\Gamma,
\{\phi\}$. Analogously to formulas, we write $\pvs{\Gamma}$ to
denote the set of all propositional variables occurring as subformulas in
formulas in $\Gamma$. For $p \in \V$, we define $\Gamma_p \isdef
\Gamma\setminus\{p\}$ for any multiset~$\Gamma$.

In the intuitionistic setting we use the following notation~$\openboxes$ on formulas: 
\[ \openboxes\psi \isdef \begin{cases} \phi &\text{ if } \psi = \Box \phi
\text{ for some formula } \phi, \\ \psi &\text{ otherwise.} \end{cases} \]
This notation is naturally overloaded to also apply to (multi)sets of formulas:
$\openboxes \Gamma \isdef \{\openboxes \phi\ |\ \phi \in \Gamma\}$.

\begin{figure}[t]
  \centering
\begin{tabular}{c@{\hspace{1cm}}c@{\hspace{1cm}}c}
$
\begin{prooftree}
\hypo{}
\infer1[\ensuremath{\IdP}]{p,\Gamma\Ra \Delta,p}
\end{prooftree}
$
&
$
\begin{prooftree}
\hypo{}
\infer1[\ensuremath{\botL}]{\bot, \Gamma\Ra \Delta}
\end{prooftree}
$

\\[1.5em]

$
\begin{prooftree}
\hypo{\Gamma\Ra \Delta, \phi}
\hypo{\psi,\Gamma\Ra \Delta}
\infer2[\ensuremath{\impL}]{\phi\rightarrow\psi,\Gamma\Ra \Delta}
\end{prooftree}
$
&
$
\begin{prooftree}
\hypo{\phi,\Gamma\Ra \Delta,\psi}
\infer1[\ensuremath{\impR}]{\Gamma\Ra \Delta,\phi\rightarrow\psi}
\end{prooftree}
$

\\[1.5em]

$
\begin{prooftree}
    \hypo{\Gamma\Ra\psi}
    \infer1[\ensuremath{\KR}]{\Phi, \Box\Gamma\Ra \Box\psi,\Delta}
\end{prooftree}
$
&
$
\begin{prooftree}
    \hypo{\Gamma, \Box\Gamma,\Box\psi\Ra\psi}
    \infer1[\ensuremath{\GLR}]{\Phi,\Box \Gamma\Ra \Box\psi,\Delta}
\end{prooftree}
$

\end{tabular}
  \caption{Classical sequent rules. Here, $\Phi$ and $\Psi$ do not contain boxed formulae.}
  \label{fig:seq-k}
\end{figure}

Now we define the sequent calculi that we use throughout the paper. The sequent
calculus $\KS$ consists of two \emph{initial}  rules $\IdP$ and $\botL$, 
left and right implication rules $\impR$ and $\impL$, and the modal rule $\KR$;
all are displayed in Figure~\ref{fig:seq-k}. The sequent calculus $\GLS$ is the
variant of the calculus $\KS$ in which the rule $\KR$ is replaced by the rule
$\GLR$ in Figure~\ref{fig:seq-k}. The sequent calculus $\KS$ is well-known to
be sound and complete for~$\K$, and $\GLS$ is sound and complete for~$\GL$
\cite{SamVal1982}. In the rule $\GLR$, the formula $\Box \psi$ is called the
\textit{diagonal} formula. We denote by $\KRprems{s}$ the multiset of all
possible $\KR$-premises for a given sequent~$s$, and by $\GLRprems{s}$ the
multiset of all $\GLR$-premises for $s$.

For $\iSL$, we work with the calculus $\GfouriSLt$ from~\cite{Shi23}, which was
specifically designed with the aim to prove uniform interpolation for $\iSL$.
The calculus is an extension of the calculus $\GfouriP$ for $\IL$~\cite{Dyc92}.
We show the calculus $\GfouriSLt$ in Figure~\ref{fig:iseq-pc}, using the
$\openboxes$ operator to rephrase its definition slightly compared
to~\cite{Shi23}.

For every sequent calculus
$\mathsf{S}$, we denote by $\vdash_{\mathsf{S}}$  the set of sequents that are
derivable using the rules in $\mathsf{S}$. For a sequent $\Gamma \To \Delta$,
we then write $\vdash_{\mathsf{S}} \Gamma \To \Delta$ to mean that $\Gamma \To
\Delta$ is an element of the set $\vdash_{\mathsf{S}}$.

The crucial fact for proving uniform interpolation is that each of the three
calculi $\KS$, $\GLS$, and $\GfouriSLt$ has a \textit{complete} and
\textit{terminating} backward proof search strategy, which may only depend on a
\textit{local} loop-check. \textit{Completeness} means that the strategy finds
a proof for any sequent provable in the calculus. \textit{Termination} means
that the strategy always ends in a finite proof search tree. By a
\textit{local} loop-check we mean: the criterion for deciding whether or not to
stop the proof search for a given sequent only depends on the sequent itself,
and does not depend on other sequents, encountered earlier by the
proof search strategy. Termination for $\KS$, $\GLS$, and $\GfouriSLt$ is
discussed in detail in Sections~\ref{sec:termination_KS},
\ref{sec:termination_GLS} and \ref{sec:termination_iSL} respectively.

\begin{figure}[t]
\centering
{\small
$\begin{prooftree}
\infer0[$\scriptstyle\rule{$\bot L$}$]{\bot, \Gamma\Ra\chi}
\end{prooftree}$
\quad
$\begin{prooftree}
\infer0[$\scriptstyle\rule{IdP}$]{\Gamma,p\Ra p}
\end{prooftree}$
\quad
$\begin{prooftree}
\hypo{\Gamma,\varphi,\psi\Ra\chi}
\infer1[$\scriptstyle\rule{$\land L$}$]{\Gamma,\varphi\land\psi\Ra\chi}
\end{prooftree}$
\quad
$\begin{prooftree}
\hypo{\Gamma\Ra\varphi}
\hypo{\Gamma\Ra\psi}
\infer2[$\scriptstyle\rule{$\land$R}$]{\Gamma\Ra\varphi\land\psi}
\end{prooftree}$
\\[0.5cm]
$\begin{prooftree}
\hypo{\Gamma,\varphi\Ra\chi}
\hypo{\Gamma,\psi\Ra\chi}
\infer2[$\scriptstyle\rule{$\lor$L}$]{\Gamma,\varphi\lor\psi\Ra\chi}
\end{prooftree}$
\quad
$\begin{prooftree}
\hypo{\Gamma\Ra\varphi_i}
\infer1[$\scriptstyle{\rule{$\lor R_i$}(i\in\{1, 2\})} $]{\Gamma\Ra\varphi_1\lor\varphi_2}
\end{prooftree}$
$\begin{prooftree}
\hypo{\Gamma,\varphi\Ra\psi}
\infer1[$\scriptstyle\rule{$\rightarrow$R}$]{\Gamma\Ra \varphi\rightarrow\psi}
\end{prooftree}$
\\[0.5cm]
$\begin{prooftree}
\hypo{\Gamma,\varphi\rightarrow (\psi\rightarrow\chi)\Ra\delta}
\infer1[$\scriptstyle\rule{$\land\!\rightarrow$L}$]{\Gamma,(\varphi\land\psi)\rightarrow\chi\Ra\delta}
\end{prooftree}$
\quad
$\begin{prooftree}
\hypo{\Gamma,\varphi\rightarrow\chi,\psi\rightarrow\chi\Ra\delta}
\infer1[$\scriptstyle\rule{$\lor\!\rightarrow$L}$]{\Gamma,(\varphi\lor\psi)\rightarrow\chi\Ra\delta}
\end{prooftree}$
\\[0.5cm]
$\begin{prooftree}
\hypo{\Gamma,p,\varphi\Ra\chi}
\infer1[$\scriptstyle\rule{$p\!\rightarrow$L}$]{\Gamma,p,p\rightarrow\varphi\Ra\chi}
\end{prooftree}$
\quad
$\begin{prooftree}
\hypo{\Gamma,\psi\rightarrow\chi\Ra \varphi\rightarrow\psi}
\hypo{\Gamma,\chi\Ra\delta}
\infer2[$\scriptstyle\rule{$\rightarrow\rightarrow$L}$]{\Gamma,(\varphi\rightarrow\psi)\rightarrow\chi\Ra\delta}
\end{prooftree}$
\\[0.5cm]
\begin{prooftree}
\hypo{\openboxes Γ \disjunion □ \phi  \Ra \phi } %
\infer1[$\scriptstyle\rule{$\Box$R}$]{Γ \Ra \Box \phi} %
\end{prooftree}
\quad
\begin{prooftree}\hypo{\openboxes Γ \disjunion  □ ϕ \disjunion  \psi \Ra ϕ }
\hypo{ Γ \disjunion  \psi \Ra \chi }
\infer2[$\scriptstyle\rule{$\Box\rightarrow$L}$]{Γ \disjunion  □ ϕ \rightarrow \psi \Ra \chi} %
\end{prooftree}
}

\caption[]{The sequent calculus~$\sys{G4iSLt}$. The sequent calculus~$\sys{G4iP}$ is the restriction of~$\sys{G4iSLt}$ obtained by omitting the two rules involving $\Box$.}

  \label{fig:iseq-pc}
\end{figure}

\subsection{Uniform interpolation}
\begin{definition} 
\label{def:UIP}
A logic~$L$ has the \emph{uniform interpolation property} if, for every
$L$-formula~$\phi$  and variable~$p$, there exist 
$L$-formulas, denoted by $\forall p \phi$ and $\exists p \phi$,
satisfying the following three properties:
\begin{enumerate}
\item \emph{$p$-freeness:} \label{UIP:1} $ \pvf{\exists p \phi} \subseteq  \pvf{\phi} \setminus \{ p \}$ and $ \pvf{\forall p \phi} \subseteq  \pvf{\phi} \setminus \{ p \}$,
\item \emph{implication:} \label{UIP:2} $\vdash_L \phi \to \exists p \phi \text{ and } \vdash_L \forall p \phi \to  \phi,$ and
\item \emph{uniformity:} \label{UIP:3} for each formula~$\psi$ with  $p \notin \pvf{\psi}$: 
\begin{align*}
    \vdash_L \phi \to \psi \ &\text{ implies } \ \vdash_L \exists p \phi \to \psi,\\
    \vdash_L \psi \to \phi \ &\text{ implies } \ \vdash_L \psi \to \forall p \phi.
\end{align*}
\end{enumerate}
\end{definition}

\begin{lemma}\label{lemma_interpolant}
Both classically and intuitionistically, the formulas $\forall p (\phi \to \psi)$ and $\exists p (\phi) \to \forall p (\phi \to \psi)$ are equivalent.
\end{lemma}
\begin{proof}
  The left-to-right direction is clear. For the right-to-left direction, note
  that the formula ${\exists p \phi \to \forall p (\phi \to \psi)}$ is
  $p$-free by definition. Moreover, one easily obtains that $\exists p \phi \to
  \forall p (\phi \to \psi)$ implies $\phi \to \psi$, using the implication
  rules and the implication properties of $\exists p$ and $\forall p$. 
  Now uniformity ensures that $\exists p \phi \to
  \forall p (\phi \to \psi)$ implies $\forall p (\phi\to\psi)$.\qed
\end{proof}

To show uniform interpolation of the logics in the paper, we employ a standard proof-theoretic approach via the sequent calculi. The following definition merges the well-known definitions for intuitionistic logic from~\cite{Pit92} and classical modal logic from~\cite{Bil06}.

\begin{definition}
\label{def SUIP}
A set of provable sequents, denoted~$\vdash$, has the \emph{uniform
interpolation property} if, for any sequent $\Gamma \To \Delta$ and variable
$p$, there exist modal formulas $\Ep{\Gamma}$ and $\Ap{\Gamma \To \Delta}$ such
that the following three properties hold: 
\begin{enumerate}
    \item \emph{$p$-freeness:} (a) $\pvf{\Ep{\Gamma}} \subseteq \pvs{\Gamma} \setminus \{ p \}$ and (b) $\pvf{\Ap{\Gamma \To \Delta}} \subseteq \pvs{\Gamma, \Delta} \setminus \{ p \}$, 
    \item \emph{implication:} (a) $\vdash \Gamma \To \Ep{\Gamma}$ and (b) $\vdash \Gamma, \Ap{\Gamma \To \Delta} \To \Delta$, and
    \item \emph{uniformity:} for any finite multisets of formulas $\Pi$ and
        $\Sigma$ such that $p \notin \pvs{\Pi,\Sigma}$, if it holds that
        $\vdash \Pi, \Gamma \To \Delta,\Sigma$, then it also holds that:  
        \begin{align*}
       &\text{(a)}\vdash \Pi, \Ep{\Gamma} \To \Delta, \Sigma \text{ if } p \notin \pvf{\Delta}, \text{ and}\\
       &\text{(b)} \vdash \Pi, \Ep{\Gamma} \To \Ap{\Gamma \To \Delta}, \Sigma.
    \end{align*} 
\end{enumerate}
In the intuitionistic setting, we require $\Delta$ to be a singleton and $\Sigma$ to be empty. 

In this paper, we say that a sequent calculus $\sf{S}$ has \emph{uniform interpolation} if $\vdash_{\sf{S}}$ has the uniform interpolation property.
\end{definition}

We provide some observations and facts in the following remarks.

\begin{remark}
When proving uniform interpolation in the classical setting, we prove a stronger statement in clause (b) of uniformity:
\begin{align*}
       &\text{(b)} \vdash \Pi \To \Ap{\Gamma \To \Delta}, \Sigma
    \end{align*} 
where we omit the occurrence of $\Ep{\Gamma}$ on the left-hand side of the sequent. In fact, now we can take $\Ep{\Gamma} := \neg \Ap{\Gamma \To \emptyset}$ and we only have to consider clauses (b) in every property of Definition~\ref{def SUIP} as in \cite{Bil06}. This will be the route taken in this paper for $\KS$ and $\GLS$.
\end{remark}

\begin{remark}\label{remark2}
It is well-known that the uniform interpolation property for a sequent calculus
results in the uniform interpolation property for its corresponding logic
\cite{Pit92,Bil07}.  Both classically and intuitionistically, we can define
$\forall p \phi \isdef \Ap{\emptyset \To \phi}$. In classical modal logic, we
can define  $\exists p \phi$ as its dual, i.e., $\exists p \phi \isdef \neg \forall p
(\neg \phi)$. For intuitionistic modal logic, we define $\exists p \phi \isdef
\Ep{\{\phi\}}$. One may then show that, for these definitions of $\forall p$
and $\exists p$, the three properties from Definition
\ref{def:UIP} follow from those in Definition~\ref{def SUIP}, where, in the
intuitionistic case, one needs to use the fact that $\Ep{\emptyset} = \top$. 
\end{remark}

\begin{remark}\label{remark:3}
In the sequel of the paper we explicitly construct operators $\Ap{\cdot}$ (and also $\Ep{\cdot}$ in the intuitionistic case)
using the terminating sequent calculi for the logics. These operators have the following properties which could be viewed as Remark~\ref{remark2} applied to sequents instead of formulas. In both the classical and intuitionistic setting, $\Ep{\Gamma}$ serves
as the formula $\exists p(\bigwedge \Gamma)$. In the classical case, the
formula $\Ap{\Gamma \To \Delta}$ will be equivalent to $\forall p (\bigwedge
\Gamma \to \bigvee \Delta)$. However, intuitionistically, $\Ap{\Gamma \To
\phi}$ is not equivalent to $\forall p (\bigwedge \Gamma \to \phi)$, but it is computed as 
$\Ep{\Gamma} \to \Ap{\Gamma \To \phi}$. The latter does not contradict Remark~\ref{remark2} by Lemma~\ref{lemma_interpolant}. See also Remark~5 in~\cite{Pit92}. 
\end{remark}

\section{Basic modal logic $\K$}
\label{sec:K}

We start our investigations on uniform interpolation for provability logics by showcasing a simple example: the modal logic $\K$. 
We follow the strategy in~\cite{Bil06} using calculus $\KS$ and provide a formalisation in Coq. 

\subsection{Termination of the sequent calculus $\KS$}\label{sec:termination_KS}

To compute the uniform interpolants for sequent calculus $\KS$, we provide a complete and terminating proof search strategy for it.
For this, we define some useful notions for sequents $\Gamma \To \Delta$. 
The \emph{size} of $\Gamma \To \Delta$ is the total number of symbols in the multiset $\Gamma, \Delta$. We call a sequent \emph{critical} if there is no formula of the form $\phi \to \psi$ in $\Gamma, \Delta$, and we call a critical sequent \emph{initial} if either $\bot \in \Gamma$ or $\Gamma \cap \Delta \cap \V \neq \emptyset$, that is, if the sequent $\Gamma \To \Delta$ can be proved with an initial rule.

A complete and terminating strategy for proof search in $\KS$ can easily be
defined in three steps, as follows.
Given a sequent, we first saturate it by maximally iterating applications of the rules $\impL$ and $\impR$.  This step computes a finite multiset $\canopy{s}$ of critical sequents, called the \emph{canopy} of $s$.
Note that, if $s$ is not critical, then all sequents in $\canopy{s}$ have strictly smaller size than~$s$.
Second, we try to apply the rules $\IdP$ and $\botL$, and close any branches where we have an initial sequent.
Third, we try to apply the rule $\KR$ on any remaining sequents which are not initial.
Since the size of sequents decreases during the execution of this
strategy as long as sequents are not initial, this strategy clearly terminates.

\subsection{Uniform interpolation for $\KS$}

\begin{definition}[\coqdoc{K.Interpolation.UIK_braga.html\#GUI}]\label{UIKdef}
Let $p\in \V$ be a variable and $s=(\Gamma, □\Gamma' \To \Delta)$ a sequent,
where no $\phi\in\Gamma$ is a boxed formula. We define $\UIK{p}{s}$ recursively, as follows: 
\begin{center}
\renewcommand{\arraystretch}{1.5}
\begin{tabular}{|@{\hspace{5pt}} c@{\hspace{5pt}}  @{\hspace{5pt}} l @{\hspace{5pt}} | @{\hspace{5pt}} l @{\hspace{5pt}} |}
    \hline 
     & \emph{if ...} & ... \emph{then} $\UIK{p}{s}$ \emph{equals:} \\
    \hline  
    $(\symUIK1)$  & $s$ \emph{is empty} & $\bot$ \\
    $(\symUIK2)$  & $s$ \emph{is not critical} & $\bigwedge\limits_{s'\in\canopy{s}} \UIK{p}{s'}$\\
    $(\symUIK3)$  & $s$ \emph{is initial} & $\top$\\
    $(\symUIK4)$  & \emph{none of the above} & 
    $\bigvee\limits_{q\in\Delta_p} q \lor \bigvee\limits_{r\in\Gamma_p}\neg r
	\lor \bigvee\limits_{s'\in \KRprems{s}}\Box \UIK{p}{s'}
	\lor \Diamond \UIK{p}{\Gamma'\Ra}$\\
    \hline
\end{tabular}
\end{center}
\end{definition}
Termination of this function is proved by an induction on the size of sequents.
This definition mirrors the termination of the proof search strategy for $\KS$.
The first case corresponds to a default where the sequent bares no content.
The remaining cases obviously correspond to steps of the strategy: $(\symUIK2)$ postpones the computation of the interpolant to the sequents in the canopy via recursive calls; $(\symUIK3)$ checks for initiality; $(\symUIK4)$ is the case where we apply $\KR$. As this last case is the most complex, we motivate that definition in more detail now.
 
Because an application of the $\KR$ rule on a sequent $s$ deletes the non-boxed
formulas in $s$, we need to first record all these formulas in $\UIK{p}{s}$:
this is the role of the first two disjuncts,
$\bigvee\limits_{q\in\Delta_p} q$ and $\bigvee\limits_{r\in\Gamma_p}\neg r$,
which notably discard all occurrences of variable $p$. 
The third disjunct, $\bigvee\limits_{s'\in \KRprems{s}}\Box \UIK{p}{s'}$, contains
recursive calls on all $\KR$-premises of $s$, and prefixes them with a $\Box$ to reflect the logical strength of the rule.
The last disjunct $\Diamond\UIK{p}{\Gamma'\Ra}$ is needed to obtain the
uniformity from Definition~\ref{def SUIP}. 
It considers the possibility that our sequent
$s=(\Gamma,\Box\Gamma'\Ra\Delta)$ becomes provable once the context is extended,
i.e., that a sequent of the
form~$\Phi,\Box\Phi',\Gamma,\Box\Gamma'\Ra\Delta,\Delta'$ is provable.
In a proof of the latter, suppose that the last rule applied was $\KR$,
triggered by a formula $\Box\phi$ in~$\Delta'$. 
In the premise $\Phi',\Gamma'\Ra\phi$ of that application, what remains of our
sequent $\Gamma,\Box\Gamma'\Ra\Delta$ is the sequent $\Gamma'\Ra$, on which we
then perform the recursive call $\UIK{p}{\Gamma'\Ra}$. 
So, the last disjunct uses a $\Diamond$ to record the possibility for
a ``step aside'' of the proof search tree, by
considering a recursive call on what remains of $s$ through a $\KR$ application
in an extended context.

The complexity of the function $\symUIK$ lies in its recursive calls on
\emph{multisets} of sequents, and in the use of the canopy function which
contains similar recursive calls. Since only computable functions can be
defined in Coq, termination needs to be proved whenever Coq cannot
automatically derive it. In order to formalise our two functions in Coq, we
synchronously need to define them and convince Coq that all recursive calls are
justified, by exhibiting a quantity which decreases along a well-founded order. Because of the complex recursive calls of our two functions,
the traditional pen-and-paper definition of such an order is rather intricate
to formalise, involving a well-founded order on multi-sets, cf.
\cite[Section~3]{Fer23}. To circumvent this difficulty in our formalisation of Definition~\ref{UIKdef} (\coqdoc{K.Interpolation.UIK_braga.html\#GUI}), we use the Braga
method~\cite{LarMon21} of Larchey-Wendling and Monin, which separates the
definition of the function from the termination proof. More precisely, using
this method we can first define a function as a relation which captures the
\emph{computational graph} of the function, and then prove that this relation
is indeed functional and terminates. While this method was initially designed to capture partial functions in Coq, we here apply this method to the definition of $\symUIK$ and the canopy. This allows us to separate the concerns of defining these functions and proving that the definition terminates.

Given that $\symUIK$ is connected to the proof search tree, and its definition
tailored to satisfy the three correctness properties for uniform interpolants,
we can now prove the correctness of the definition, and formalise it in Coq.

\begin{theorem}\label{SUIPKS}
The sequent calculus $\KS$ has the uniform interpolation property.
\end{theorem}

\begin{proof}
We have formalised in the Coq proof assistant the proof from~\cite{Bil06} with
no major changes. We have to check the three properties from
Definition~\ref{def SUIP}, i.e., $p$-freeness, implication, and uniformity. It
is evident that $\Ap{s}$ is $p$-free for every sequent~$s$, as the computations
in $\symUIK$ all make sure to discard $p$ whenever propositional variables are
recorded (\coqdoc{K.Interpolation.UIK_UIOne.html\#UI_One}). Second, as
$\UIK{p}{\Gamma\Ra\Delta}$ follows closely the proof search tree of
$\Gamma\Ra\Delta$, we obtain rather straightforwardly that
$\UIK{p}{\Gamma\Ra\Delta},\Gamma\Ra\Delta$ is provable
(\coqdoc{K.Interpolation.UIK_UITwo.html\#UI_Two}), hence proving the
implication property. Finally, we make a crucial use of the disjunct
$\Diamond\UIK{p}{\Gamma\Ra}$ of the case $(\symUIK4)$ in the proof of
uniformity (\coqdoc{K.Interpolation.UIK_UIThree.html\#UI_Three}).\qed
\end{proof}

\section{Classical provability logic $\GL$}
\label{sec:GL}

We now shift our focus to the logic $\GL$. We will first provide a complete and terminating strategy for $\GLS$. 
Then, in order to construct uniform interpolants for $\GL$, we take inspiration from \cite{Bil22},
but we modify the definition given there in order to fix an incompleteness in the correctness proof.

\subsection{Terminating strategy for sequent calculus $\GLS$}\label{sec:termination_GLS}
In the rule $\GLR$, the multiset $\Box\Gamma$ on the left of the premise is preserved,
while the diagonal formula $\Box \psi$ moves diagonally from the left to the right when moving from premise to conclusion.
These features are known to be an obstacle to the termination of a strategy for $\GLS$, 
which can be overcome by a local loop-check.
Consider the following rule, labelled $\IdB$ for `Identity Box'.
\begin{center}
\begin{tabular}{c}
$
\begin{prooftree}
\hypo{}
\infer1[\ensuremath{\IdB}]{\Box\phi,\Gamma\Ra \Delta,\Box\phi}
\end{prooftree}
$
\end{tabular}
\end{center}
Our proof search strategy for $\GLS$ extends the one for $\KS$: 
first apply $\impL$ and $\impR$,
then the initial rules $\IdP$, $\botL$ and $\IdB$, and finally the rule $\GLR$.
When following this strategy, any application of the rule $\GLR$ 
is such that its conclusion is critical but not initial, where our definition of \emph{initial} sequent
now also includes sequents that allow for an application of $\IdB$.
Note a subtlety of our strategy: while $\IdB$ is not a rule of $\GLS$ its presence in our strategy is justified by its \emph{admissibility} \cite{GorRamShi21}, ensuring the completeness of this strategy.

To show termination, we define a measure on sequents which decreases, in a well-founded order, 
as we move upwards by applying rules according to the proof strategy.
Given a sequent $\Gamma\Ra\Delta$, its measure $\GLseqmeas{\Gamma\Ra\Delta}$ is a pair of natural numbers $(\impocc{\Gamma\Ra\Delta}\, ,\,\numusablebox{\Gamma\Ra\Delta})$, 
where the first component is the number of occurrences of the symbol $\rightarrow$ in $\Gamma\Ra\Delta$ and the 
second component is what we call the \emph{number of usable boxes}, $\numusablebox{\Gamma \Ra \Delta}$, 
defined as the cardinal of the set 
$\{\Box\phi\mid \Box\phi\in \sbf{\Gamma\cup\Delta}\}\setminus\{\Box\phi\mid \Box\phi \in\Gamma\}$. 
The idea is that $\symnumusablebox$ counts 
the number of boxed formulas of a sequent $\Gamma\Ra\Delta$ which might later become the diagonal 
formula of an instance of $\GLR$ in a derivation of this sequent, when following the proof search strategy. 
To show termination of our strategy via $\Theta$, we use the lexicographic order $\lexnat$ on pairs of natural 
numbers, noting that, for any $\GLS$ rule with conclusion $s$ and any premise $s'$ of that rule, 
we have $\GLseqmeas{s'}\lexnat\GLseqmeas{s}$.

\subsection{Computing uniform interpolants for $\GLS$}
We now replicate the argument for $\K$ for $\GL$, using the sequent calculus $\GLS$ and the terminating and complete proof search strategy for it.
A first try would be to use the modified notion of initiality, and to change the function $\symUIK$ into a function $\symUIGL$ by exchanging the rule $(\symUIK4)$ for a similar rule that 
follows the rule $\GLR$ instead of $\KR$.
However, this approach leads to a termination problem in the fourth case of the definition of the function, as 
was noticed in \cite{Bil06}, and as we briefly explain now.
In this case $\Gamma, \Box\Gamma'\Ra\Delta$ is critical, not empty and not initial, 
so we would require a recursive call of the function on $\Gamma'\Box\Gamma'\Ra$ in the last disjunct. 
However, this recursive call could fail to terminate, as we do not have in general that 
$\GLseqmeas{\Gamma', \Box\Gamma'\Ra}\lexnat\GLseqmeas{\Gamma, \Box\Gamma'\Ra\Delta}$.
To address this problem, \cite{Bil06} used 
an auxiliary function $\symN$ in the definition of $\symUIGL$ for $\GL$. 

We recall the definition of the function $\mathsf{N}$ as given in \cite{Bil22} in Figure~\ref{fig:N_GL}; 
in Definition~\ref{UIGLdef} below, we will modify this table to obtain 
a mutually recursive definition of the function $\symUIGL$. 
Given the function $\mathsf{N}$, the idea is, then, 
to replace the rule ($\symUIK4$) in Definition~\ref{UIKdef} by a rule which says
that, if $s = (\Gamma, \Box \Gamma' \Ra \Delta)$ and $s$ is critical, not empty, and not initial, then
$\UIGL{p}{s}$ equals
\begin{equation}\tag{$\symUIGL4$}\label{symUIGL4}
    \bigvee\limits_{q\in\Delta_p} q \lor \bigvee\limits_{r\in\Gamma_p}\neg r
	\lor \bigvee\limits_{s'\in \GLRprems{s}}\Box \UIGL{p}{s'}
	\lor \Diamond \bigwedge\limits_{t\in\canopy{\Gamma', \Box\Gamma'\Ra}} \N{s,t}\ . 
 \end{equation}

\begin{figure}[t]
\begin{center}
\setlength{\extrarowheight}{4pt}
\begin{tabular}{|@{\hspace{5pt}} c@{\hspace{5pt}}  @{\hspace{5pt}} l @{\hspace{5pt}} | @{\hspace{5pt}} l @{\hspace{5pt}} |}

\hline
  & {if \dots} & \dots then $\N{s,t}$ equals\\
 \hline 
    $(\symN1)$  & $t$ is initial & $\top$\\
    $(\symN2)$ & $t$ is not initial and $\numusablebox{t}<\numusablebox{s}$ & $\UIGL{p}{t}$ \\
    $(\symN3)$  & otherwise & $\bigvee\limits_{q\in\Pi_p} q \lor \bigvee\limits_{r\in\Sigma_p}\neg r
	\lor \bigvee\limits_{t'\in \GLRprems{t}}\Box\UIGL{p}{t'}$\\
    \hline
\end{tabular}
\caption{Definition of function $\N{\cdot,\cdot}$ from \cite{Bil06}, where $t = (\Sigma \Ra \Pi)$.}
\label{fig:N_GL}
\end{center}
\end{figure}
Here, in the 
last disjunct of $(\symUIGL4)$, 
we apply the function $\symN$ to all elements of the canopy of the sequent $\Gamma', \Box\Gamma'\Ra$, which is exactly
what remains of the sequent $s$ after applying $\GLR$ upwards.
The purpose of the function $\symN$ is to attempt another unfolding of $\symUIGL$ in the canopy of $\Gamma',\Box\Gamma'\Ra$. 
Indeed, the definition of $\symN$ first checks whether any recursive call is necessary via the initiality check in $(\symN1)$, and then proceeds in $(\symN2)$ to recursively call $\symUIGL$ 
if we are ensured that $\Theta$ decreases via the first component, 
or goes to $(\symN3)$ if there is no such decrease.
Notice that, in this last case, the definition of $\symN$ is a truncation of $(\symUIGL4)$,
which omits the problematic last disjunct, as it cannot be guaranteed to decrease in the recursion. 
The termination of $\symUIGL$ is obviously ensured by definition. However,
the correctness is no longer obvious, due to the truncation in the rule $(\symN3)$.
The key insight for proving the correctness is the following \emph{fixed point} equivalence~\cite{Bil22}
which is valid in $\GL$:
$$\Diamond\left(\bigwedge\limits_{i}\left[\alpha_i\lor\Diamond\left(\bigwedge\limits_{i}\alpha_i\land\beta\right)\right]\land\beta\right)
\hspace{1cm}
\leftrightarrow
\hspace{1cm}
\Diamond\left(\bigwedge\limits_{i}\alpha_i\land\beta\right)\ . $$
This equivalence can be used to prove that the diamond disjunct from the rule
$(\symUIGL4)$ may be omitted in the rule $(\symN3)$. 
In order to make this work formally, one needs the following equivalence to be derivable in $\GLS$:
\begin{equation}\label{keyeq}
\Diamond \bigwedge\limits_{s'\in\canopy{\Gamma', \Box\Gamma'\Ra}} \N{s,s'} \quad \leftrightarrow \quad \Diamond\UIGL{p}{\Gamma',\Box\Gamma'\Ra}\ .
\end{equation}
Assuming this equivalence, one can show that the uniform interpolation property holds for $\GLS$.
To justify (\ref{keyeq}), \cite{Bil22} relies on another equivalence between 
two formulas $\N{s,t_1}$ and $\N{s,t_2}$, where $t_i = \Gamma_i, \Box\Gamma_i \Ra$ for $i = 1,2$, where the 
multisets $\Gamma_1$ and $\Gamma_2$ are known to be equal only when considered \emph{as sets}, i.e., not counting multiplicities. 
This equivalence is not formally proved, but only ``observe[d]" \cite[p.~17]{Bil22}. 
Since the sequents $t_1$ and $t_2$ are 
\emph{identical modulo contraction}, and contraction is an admissible rule in $\GLS$, 
this sounds reasonable, but we were unable to formally derive this equivalence, 
even after consulting with the author of \cite{Bil22}.

The difficulty in formally proving the observation 
primarily lies in the fact that the function $\symN$ includes computations of the canopy of our two sequents $t_1$ and $t_2$. 
However, the canopies of two sequents can vastly differ, even if they are identical modulo contraction. 
We give a minimal example of such a situation in Figure~\ref{fig:ex-canopy}, where the sequents $q\Ra p$ on the right find no counterparts on the left. 
This mismatch in canopies, then, makes it hard to prove that any call to $\symUIGL$ in one canopy has a counterpart in the other canopy.

\begin{figure}
\begin{tabular}{c @{\hspace{1cm}} c}
    $\begin{prooftree}
    \chypo{\Ra p}
    \chypo{q\Ra}
    \infer2[\ensuremath{\impL}]{p\rightarrow q\Ra}
    \end{prooftree}
    $
    & 
    $
    \begin{prooftree}
    \chypo{\Ra p,p}
    \chypo{q\Ra p}
    \infer2[\ensuremath{\impL}]{p\rightarrow q\Ra p}
    \chypo{q\Ra p}
    \chypo{q,q\Ra}
    \infer2[\ensuremath{\impL}]{p\rightarrow q,q\Ra}
    \infer2[\ensuremath{\impL}]{p\rightarrow q,p\rightarrow q\Ra}
    \end{prooftree}$
\end{tabular}
\caption{Two sequents that are equivalent up to contraction, but the canopies are not.}
\label{fig:ex-canopy}
\end{figure}
In order to overcome this problem, we propose to modify the mutually recursive definition
of $\symUIGL$ and $\symN$ with respect to the one given in \cite{Bil22}:
in strategic places, we \emph{fully contract} sequents, notably before computing canopies. 
We denote by $\fc{s}$ the fully contracted version of the sequent $s$; that is, when $s = (\Gamma \Ra \Delta)$, 
$\fc{s}$ denotes the sequent $(\Gamma' \Ra \Delta')$, where $\Gamma'$ and $\Delta'$ are the multisets obtained
from $\Gamma$ and $\Delta$, respectively, by removing duplicates.

\begin{definition}[\coqdoc{GL.Interpolation.UIGL_braga.html\#GUI}]\label{UIGLdef}
Let $p\in \V$ be a variable. We define $\symUIGL$ and $\symN_p$ by a mutual recursion, as follows.
Let $s=(\Gamma, □\Gamma' \To \Delta)$ be a sequent, where no $\phi\in\Gamma$ is a boxed formula.
If $s$ is empty or initial, then $\UIGL{p}{s}$ equals $\UIK{p}{s}$, and
\begin{center}
\setlength{\extrarowheight}{4pt}
\begin{tabular}{|@{\hspace{5pt}} c@{\hspace{5pt}}  @{\hspace{5pt}} l @{\hspace{5pt}} | @{\hspace{5pt}} l @{\hspace{5pt}} |}
    \hline
    & \emph{if \dots} & \emph{\dots then } $\UIGL{p}{s}$ \emph{equals}\\
    \hline
    $(\symUIGL2)$ & $s$ \emph{is not critical} & $\bigwedge\limits_{s'\in\canopy{\fc{s}}} \UIGL{p}{s'}$  \\
    $(\symUIGL4)$ & \emph{otherwise} & 
    \begin{tabular}{@{}r@{}}
    $\bigvee\limits_{q\in\Delta_p} q \lor \bigvee\limits_{r\in\Gamma_p}\neg r
	\lor \bigvee\limits_{s'\in \GLRprems{\fc{s}}}\Box \UIGL{p}{s'}$\\
    $
	\lor \Diamond \bigwedge\limits_{t\in\canopy{\fc{\Gamma, \Box\Gamma\Ra}}} \N{s,t}$
    \end{tabular}\\
    \hline
\end{tabular}
\end{center}
Let $t = (\Sigma \Ra \Pi)$ be a sequent.
We also define \emph{(}\coqdoc{GL.Interpolation.UIGL_braga.html\#GN}\emph{)} the formula $\N{s,t}$ as 
in Figure~\ref{fig:N_GL}, but replacing the formula in the last row of the table with:
$$\bigvee\limits_{q\in\Pi_p} q \lor \bigvee\limits_{r\in\Sigma_p}\neg r
	\lor \bigvee\limits_{t'\in \GLRprems{\fc{t}}}\Box\UIGL{p}{t'}\ , \vspace{-0.4em} $$
where we note that the last disjunction is indexed by $\GLRprems{\fc{t}}$ instead of $\GLRprems{t}$.
\end{definition}

With this new definition, we obtain a proof of correctness of the equivalence~(\ref{keyeq}),
as we always fully contract sequents before computing their canopies.
In our formalisation of Definition~\ref{UIGLdef}, 
we again made use of the Braga method already described in Section~\ref{sec:K}.

\subsection{Syntactic correctness proof}

\begin{theorem}\label{SUIPGLS}
The sequent calculus $\GLS$ has the uniform interpolation property.
\end{theorem}

\begin{proof}
We refer to the formalised proofs of the first
(\coqdoc{GL.Interpolation.UIGL_UIOne.html\#UI_One}), second
(\coqdoc{GL.Interpolation.UIGL_UITwo.html\#UI_Two}) and third
(\coqdoc{GL.Interpolation.UIGL_UIThree.html\#UI_Three}) property.\qed
\end{proof}

\newpage
\section{Intuitionistic strong Löb $\iSL$}
\label{sec:iSL}

The aim of this section is to give a sequent-based proof of the uniform
interpolation property for intuitionistic strong Löb logic, $\iSL$. We will
simultaneously explain the proof method of this new result, and report on our
mechanisation of the definition of the propositional quantifiers in Coq. 
The work in this section builds on an
earlier formalisation~\cite{Fer23} of Pitts' theorem~\cite{Pit92} that uniform
interpolation holds for $\IL$. In order to make the explanation below for
$\iSL$ understandable, we first briefly review some important points of that
work.
We subsequently explain how to
extend that definition to deal with the modality of the logic $\iSL$, and how
the correctness proof can be extended to work for that logic.

As for the classical modal logics considered above,
the definitions of the propositional quantifiers $\Ap{\cdot}$ and $\Ep{\cdot}$
for $\IL$ are guided by 
the terminating sequent calculus, $\GfouriP$ (see Figure~\ref{fig:iseq-pc}).
In~\cite{Pit92,Fer23}, $\Ap{\cdot}$ and $\Ep{\cdot}$ are defined for $\GfouriP$ as follows. Based on the rows $(\EpIPC0)$-$(\EpIPC8)$ and $(\ApIPC1)$-$(\ApIPC13)$ in Figure~\ref{fig:UIP_IPC}, the sets $\callAp{\Gamma \To \phi}$ and $\callEp{\Gamma}$ are defined by pattern matching. Based on this we define,
\begin{align} \label{IPC_normalform}
\ApIPC(\Gamma \To \phi) := \bigvee \callAp{\Gamma \To \phi} \ \ \text{ and } \ \ \EpIPC(\Gamma) := \bigwedge \callEp{\Gamma}.
\end{align}

\begin{theorem}\label{thm:pitts}
The sequent calculus for $\IL$ has the uniform interpolation property.
\end{theorem}

 \subsection{Termination of sequent calculus  $\GfouriSLt$} \label{sec:termination_iSL}

The calculus $\GfouriSLt$ has already been shown to be terminating
\cite{Shi23}, 
but we find it convenient to provide a different termination ordering here, which is closer to, and
compatible with, the termination ordering used by Pitts in the context of the
sequent calculus $\GfouriP$, also see~\cite{Dyc92,DycNeg00}. In particular,
this lets us re-use some earlier Coq engineering
work~\cite[Thm.~3.3]{Fer23} that was needed to be able to apply the theorem of
Dershowitz and Manna~\cite{DerMan} 
that the natural order on the set of multisets of well-founded order is
again well-founded.
The \emph{weight} of a formula is inductively defined, by adding a given weight for each symbol:
$\bot, \Box, \rightarrow$ and variables count for $1$, $\wedge$ for $2$ and $\vee$ for $3$. 
This naturally defines a well-founded strict preorder on the set of formulas:
$\phi \prec_f \psi$ iff $\weight(\phi) < \weight(\psi)$.

In \cite{Dyc92}, the preorder on sequents used  to prove the
termination of $\GfouriP$ is the \emph{Dershowitz-Manna} ordering on multisets
induced by this ordering on formulas:
$\Gamma \Ra \phi \prec \Delta \Ra \psi$ if the multiset $\Gamma \disjunion
\phi$ is smaller than the multiset $\Delta \disjunion \psi$.
However, the \boxR-rule of $\GfouriSLt$ is not always compatible with this ordering.
Indeed, with $\Gamma = \emptyset$ and $\phi = \bot$, note that $\{\Box \bot, \bot\}
\not\prec \{\Box \bot\}$.
The reason is that this rule both replaces a boxed formula on the right hand side
with its unboxed version, which is a strict subformula, but also moves the boxed formula
to the left-hand side.

We fix this issue by counting twice the right-hand side of the sequent in the
multiset, accounting for the fact that a formula on the right-hand side of a sequent 
might be duplicated using a $\boxR$ rule.

\begin{center}
\begin{figure}[!ht] 
\begin{tabular}{| l | l | l |}
\hline
    & $\Gamma$ matches & $\callEp{\Gamma}$ contains\\
\hline
    $(\EpIPC0)$ & $\Gamma',\bot$ & $\bot$ \\
    $(\EpIPC1)$ & $\Gamma',q$ & $\Ep{\Gamma'} \wedge q$ \\
    $(\EpIPC2)$ & $\Gamma',\psi_1 \wedge \psi_2$ & $\Ep{\Gamma', \psi_1, \psi_2}$ \\
    $(\EpIPC3)$ & $\Gamma',\psi_1 \vee \psi_2$ & $\Ep{\Gamma',\psi_1} \vee \Ep{\Gamma',\psi_2}$ \\
    $(\EpIPC4)$ & $\Gamma',(q \to \psi)$ & $q \to \Ep{\Gamma',\psi}$ \\
    $(\EpIPC5)$ & $\Gamma',p, (p \to \psi)$ & $\Ep{\Gamma',p, \psi}$ \\
    $(\EpIPC6)$ & $\Gamma',(\delta_1 \wedge \delta_2) \to \delta_3)$ & $\Ep{\Gamma',(\delta_1 \to (\delta_2 \to \delta_3))}$ \\
    $(\EpIPC7)$ & $\Gamma',(\delta_1 \vee \delta_2) \to \delta_3)$ & $\Ep{\Gamma',(\delta_1 \to \delta_3),(\delta_2 \to \delta_3))}$ \\    
    $(\EpIPC8)$ & $\Gamma',(\delta_1 \to \delta_2) \to \delta_3)$ & \begin{tabular}{@{}r@{}}
    $[\Ep{\Gamma',(\delta_2 \to \delta_3)} \to \Ap{\Gamma',(\delta_2 \to \delta_3) \To \delta_1 \to \delta_2}]$ \\
    $\to \Ep{\Gamma',\delta_3}$
    \end{tabular}
      \\
\hline  
    $(\symEpiSL9)$ & $\Gamma',\Box \delta$ & 
    $\Box \Ep{\openboxes \Gamma' \disjunion \delta}$ \\
    $(\symEpiSL10)$ & $\Gamma',(\Box \delta_1 \rightarrow \delta_2)$ & 
    \begin{tabular}{@{}r@{}}
    $□[\Ep{⊗\Gamma' \disjunion  \delta_2 \disjunion  □\delta_1}\rightarrow  \Ap{⊗\Gamma' \disjunion  \delta_2 \disjunion  □\delta_1 \Ra \delta_1}]$\\
    $ \rightarrow  \Ep{\Gamma' \disjunion  \delta_2}$
    \end{tabular}
    \\
\hline \hline
    & $s$ matches & $\callAp{s}$ contains\\
\hline
    $(\ApIPC1)$ & $\Gamma,q \To \phi$ & $\Ap{\Gamma \To \phi}$ \\
    $(\ApIPC2)$ & $\Gamma,\psi_1 \wedge \psi_2 \To \phi$ & $\Ap{\Gamma, \psi_1, \psi_2 \To \phi}$ \\
    $(\ApIPC3)$ & $\Gamma,\psi_1 \vee \psi_2 \To \phi$ & \begin{tabular}{@{}r@{}}
    $[\Ep{\Gamma,\psi_1} \to \Ap{\Gamma,\psi_1 \To \phi}] \wedge$\\
    $[\Ep{\Gamma,\psi_2} \to \Ap{\Gamma,\psi_2 \To \phi}]$
    \end{tabular}\\
    $(\ApIPC4)$ & $\Gamma,(q \to \psi) \To \phi$ & $q \wedge \Ap{\Gamma,\psi \To \phi}$ \\
    $(\ApIPC5)$ & $\Gamma,p, (p \to \psi) \To \phi$ & $\Ap{\Gamma,\psi \To \phi}$ \\
    $(\ApIPC6)$ & $\Gamma,(\delta_1 \wedge \delta_2) \to \delta_3) \To \phi$ & $\Ap{\Gamma,(\delta_1 \to (\delta_2 \to \delta_3)) \To \phi}$ \\
    $(\ApIPC7)$ & $\Gamma,(\delta_1 \vee \delta_2) \to \delta_3) \To \phi$ & $\Ap{\Gamma,(\delta_1 \to \delta_3),(\delta_2 \to \delta_3)) \To \phi}$ \\    
    $(\ApIPC8)$ & $\Gamma,(\delta_1 \to \delta_2) \to \delta_3) \To \phi$ & 
    \begin{tabular}{@{}r@{}}
    $[\Ep{\Gamma,(\delta_2 \to \delta_3)} \to \Ap{\Gamma,(\delta_2 \to \delta_3) \To \delta_1 \to \delta_2}]$\\
    $\wedge~\Ap{\Gamma,\delta_3 \To \phi}$
    \end{tabular}\\
    $(\ApIPC9)$ & $\Gamma \To q$ & $q$ \\
    $(\ApIPC10)$ & $\Gamma, p \To p$ & $\top$\\   
    $(\ApIPC11)$ & $\Gamma \To \phi_1 \wedge \phi_2$ & $\Ap{\Gamma \To \phi_1} \wedge \Ap{\Gamma \To \phi_2}$\\
    $(\ApIPC12)$ & $\Gamma \To \phi_1 \vee \phi_2$ & $\Ap{\Gamma \To \phi_1} \vee \Ap{\Gamma \To \phi_2}$  \\    
    $(\ApIPC13)$ & $\Gamma \To \phi_1 \to \phi_2$ & $\Ep{\Gamma,\phi_1} \to \Ap{\Gamma, \phi_1 \To \phi_2}$\\
\hline 
    $(\symApiSL14)$ & $\Gamma \To \Box \delta$ & $□(\Ep{⊗\Gamma \disjunion  □\delta} \rightarrow \Ap{⊗\Gamma \disjunion  □\delta\Ra \delta}).$\\ 
    $(\symApiSL15)$ & $\Gamma \disjunion  □\delta_1 \rightarrow \delta_2 \Ra \phi $ & 
    \begin{tabular}{@{}r@{}}
    $□[\Ep{⊗\Gamma \disjunion \delta_2 \disjunion  □\delta_1}\rightarrow \Ap{⊗\Gamma \disjunion  \delta_2 \disjunion  □\delta_1\Ra \delta_1}]$\\
    $ ∧ \Ap{\Gamma \disjunion  \delta_2 \Ra ϕ}$
    \end{tabular}\\
\hline
\end{tabular}
\caption{The top part of each table, i.e., $(\EpIPC0)$-$(\EpIPC8)$ and $(\ApIPC1)$-$(\ApIPC13)$ define $\Ep{\Gamma}$ and $\Ap{\Gamma \To \phi}$ for $\IL$ as defined in~\cite{Pit92}. The complete table provides definitions for $\Ep{\Gamma}$ and $\Ap{\Gamma \To \phi}$ for $\iSL$. In all clauses, $q \neq p$.}
\label{fig:UIP_IPC}
\end{figure}
\end{center}

\begin{definition}[Sequent ordering]
$\Gamma \Ra \phi \prec \Delta \Ra \psi$ whenever
$\Gamma \disjunion \phi \disjunion \phi$ is smaller than $\Delta \disjunion \psi
\disjunion \psi$ for the multiset ordering induced by $\prec_f$.
\end{definition}

The ordering is again well-founded, as follows from an application of the
Dershowitz-Manna theorem to the fact that the weight ordering on formulas is
well-founded. Also, any hypothesis of an $\GfouriSLt$ rule is
smaller than its conclusion.
This ensures the termination of proof search for $\GfouriSLt$, but we will also use this
ordering to construct the uniform interpolants.

Note that, although this order does not strictly speaking contain the original
order,  it is the case that, if two sequents were comparable 
for the original one in Pitts proof, then they still are for this modified
order.
This means that changing the definition of the ordering does not break the proof structure 
for the existing cases with no modality involved. This allows us 
to adapt the existing Coq formalisation for $\GfouriP$ at minimal cost.

\subsection{Computing uniform interpolants for  $\GfouriSLt$}
Following the same proof scheme as Pitts'
for $\IL$, we now define $\EpiSL{\Gamma}$ and
$\ApiSL{\Gamma\Ra\phi}$.
\begin{definition}
    The formulas $\EpiSL{\Gamma}$ and $\ApiSL{\Gamma\Ra\phi}$ are defined by mutual induction on the $\prec$ ordering, respectively as a conjunction of a multiset of formulas $\callEp{\Gamma}$ and as a disjunction 
 of a multiset of formulas $\callAp{\Gamma \Ra \phi}$, both defined by the rules from Figure~\ref{fig:UIP_IPC}.
\end{definition}

\begin{remark}
Our adaptation of Pitts' construction for $\IL$ to $\iSL$ adds formulas to the sets $\mathcal{E}_p$ and $\mathcal{A}_p$ 
only in the
cases where some formula in $\Delta \disjunion \theta$ contains a boxed
subformula. As a consequence, $\ApiSL{\Gamma \Ra \phi} = \ApIPC{(\Gamma \Ra \phi)}$ and $\EpiSL{\Gamma} = \EpIPC{(\Gamma)}$ whenever $\Gamma$ and $\phi$ do not contain the $\Box$ modality.
\end{remark}

\begin{remark}
Rule $(\symEpiSL9)$ can be read as adding $\Box \EpiSL{\openboxes \Gamma}$ to the set $\callEp{\Gamma}$ whenever $\Gamma$ contains at least one boxed formula (otherwise, $\openboxes \Gamma = \Gamma$ and this definition would not be well-founded). An efficient implementation of this rule should then take care not to add multiple copies of $\Box \EpiSL{\openboxes \Gamma}$, i.e. for each boxed formula in $\Gamma$.
\end{remark}

In order to prove the \textit{implication} and \textit{uniformity} properties of uniform interpolation (Definition~\ref{def SUIP}) we will first require some admissibility lemmas for $\GfouriSLt$, in particular weakening and contraction.
Note that, as for Pitts' proof for $\IL$, the admissibility of cut is not necessary here and indeed, we do not use nor prove it in our Coq mechanisation. However, since cut is in fact admissible in $\GfouriSLt$~\cite{Shi23}, 
we allow ourselves to use this fact in our `paper' explanations below.
In addition, $\iSL$ satisfies the strongness property.
\begin{lemma}[Strongness] \label{lem:strongness}
  For any formula $\phi$, $\vdash_{\iSL} \phi \Ra \Box \phi $.
\end{lemma}

However, we will actually use the following stronger, dual lemma instead, provable by induction on the proof derivation of $\vdash_{\iSL} \Delta \disjunion \phi \Ra \phi$.

\begin{lemma}\label{lem:open_boxesL}
  If $\vdash_{\iSL}\Delta \disjunion \phi \Ra \psi$ then $\vdash_{\iSL} \Delta \disjunion \openboxes \phi \Ra \psi$.
\end{lemma}

The following lemma highlights how the interpolant interacts with the $\Box$ modality and its dual $\openboxes$.
\begin{lemma}\label{lem:Ebox} For any multiset of formulas $\Delta$,
  $\vdash_{\iSL} \EpiSL{\Delta}\Ra \Box \EpiSL{\openboxes \Delta}\ .$
\end{lemma}
\begin{proof}
  If $\Delta$ contains no boxed formulas, then $\openboxes\Delta = \Delta$ and
  Lemma~\ref{lem:strongness} lets us conclude.
  Otherwise, $\Delta$ is multiset-equivalent to $\Delta' \disjunion \Box \delta$ for
  some $\Delta'$ and $\delta$.
  Then, by rule $(\symEpiSL9)$, $\EpiSL{\Delta}$ is a conjunction containing
  $\Box(\EpiSL{\openboxes \Delta'\disjunion \delta})$ which is equivalent to
  $\Box(\EpiSL{\openboxes \Delta})$ since the definition of $\EpiSL{\cdot}$ is invariant
  under multiset-equivalence.
  \qed
\end{proof}

\begin{theorem}
    The sequent calculus $\GfouriSLt$ has uniform interpolation.
\end{theorem}
\begin{proof}
  The \textit{$p$-freeness} property is easily proved (\coqdoc{ISL.PropQuantifiers.html\#EA_vars}).
    The \textit{implication} property is proved (\coqdoc{ISL.PropQuantifiers.html\#entail_correct}) by well-founded induction of $\prec$ on the sequent $\Delta \Ra ϕ$ and mostly relies on weakening.
    The proof of \textit{uniformity} (\coqdoc{ISL.PropQuantifiers.html\#pq_correct}) is by structural induction on the
    derivation of~$\vdash_{\iSL} Γ, Δ \Ra ϕ$. If the last rule is an $\IL$
    rule, then Pitts' proof of uniform interpolation for $\IL$ still applies.
    The cases for the modal rules are handled similarly, with a critical use of
    Lemmas~\ref{lem:open_boxesL}~and~\ref{lem:Ebox}.
    We postpone a detailed pen-and-paper version to a forthcoming journal publication.\qed
\end{proof}

\section{Conclusion and future work}
\label{sec:conclusion}
We have provided formalised sequent-style proofs of 
three uniform interpolation results, one well-known ($\K$), a 
second subtle ($\GL$), and a third new ($\iSL$).
One recent application of the verified implementation of uniform interpolation of $\IL$~\cite{Fer23}
was to prove non-definability results in intuitionistic logic~\cite{Koc2023}. 
We hope that the implementations given in this paper and the accompanying \textcolor{blue}{\href{https://hferee.github.io/UIML/demo.html}{online demo}} can be similarly useful in the future.

As explained in detail in Section~\ref{sec:GL}, 
our effort made in formalising the argument of \cite{Bil22} in Coq exposed an incompleteness in the paper proof,
which we were eventually able to correct.
This incompleteness would not have been discovered (nor corrected)
as quickly without the formalisation effort. 
The work in that section thus provides a further example of the usefulness of such efforts 
when subtle correctness proofs of algorithms in logic are concerned.

We leave to future work a more modular formal development of uniform interpolation proofs.
In particular, one could formalise the theoretical results of \cite{Iem2019} in order to 
obtain a general algorithm which, given as input a sufficiently well-behaved sequent calculus,
produces a verified calculation of uniform interpolants for the corresponding logic.
A further piece of evidence that such a general development might be possible is that the
generalisation from the known result for the logic $\IL$ to the new result for the logic $\iSL$ was
relatively frictionless. This shows another strength of the formalisation endeavour, allowing for an easy
experimentation with the boundaries of the formalised results.

A concrete logic that we would like to capture with our work is the intuitionistic version of $\GL$, 
often referred to as $\iGL$, for which it is an open problem whether or not uniform interpolation holds~\cite{Gie22}.

A final problem that we leave to future work is the formalisation of
the semantic approach to uniform interpolation, 
via the definability of bisimulation quantifiers, as e.g. in \cite{Vis1996,GhiZaw2002,GieJalKuz21,GieJalKuz23}.
This would allow for a comparison of the two approaches, both in terms of algorithmic complexity
and ease of formalisation.

\textbf{Acknowledgments.} We thank Marta Bílková, Dominique Larchey-Wendling, and Tadeusz Litak for fruitful discussions. This research received funding from the Agence Nationale de la Recherche, project ANR-23-CE48-0012. This work was partially supported by a UKRI Future Leaders Fellowship, ‘Structure vs Invariant in Proofs’, project reference MR/S035540/1.

\bibliographystyle{splncs04}
\bibliography{bibliography}

\end{document}